\documentclass[prl,twocolumn]{revtex4}%
\usepackage{amsfonts}
\usepackage{amsmath}
\usepackage{amssymb}
\usepackage{graphicx}%
\providecommand{\U}[1]{\protect\rule{.1in}{.1in}}

\begin{document}
\title{Intrinsic Spin Hall Effect Induced by Quantum Phase Transition in HgCdTe Quantum Wells}
\author{Wen Yang}
\author{Kai Chang}
\altaffiliation[Electronic address:]{kchang@semi.ac.cn}

\affiliation{State Key Laboratory for Superlattices and Microstructures, Institute of
Semiconductors, Chinese Academy of Sciences, P. O. Box 912, 100083, Beijing, China}
\author{Shou-Cheng Zhang}
\affiliation{Department of Physics, McCullough Building, Stanford University, Stanford, CA
94305-4045, USA}


\begin{abstract}
Spin Hall effect can be induced both by the extrinsic impurity
scattering and by the intrinsic spin-orbit coupling in the
electronic structure. The HgTe/CdTe quantum well has a quantum
phase transition where the electronic structure changes from
normal to inverted. We show that the intrinsic spin Hall effect of
the conduction band vanishes on the normal side, while it is
finite on the inverted side. This difference gives a direct
mechanism to experimentally distinguish the intrinsic spin Hall
effect from the extrinsic one.
\end{abstract}
\maketitle

Spin-polarized transport in nonmagnetic semiconductors is a
crucial ingredient for realizing spintronic
devices.\cite{Spintronics} The spin Hall effect (SHE) opens up the
promising prospect of generating spin currents in conventional
semiconductors without applying external magnetic field or
introducing ferromagnetic elements. Recently the previously
predicted extrinsic SHE (ESHE)\cite{ESHE} and the newly discovered
intrinsic SHE (ISHE)\cite{ISHE} have become one of the most
intensively studied subjects. The experimental observations of SHE
have been reported by two groups\cite{AwschalomSHE,WunderlichSHE}
in \textit{n}-type epilayers and two-dimensional electron and hole
gases, although their theoretical interpretation as extrinsic or
intrinsic are still
ambiguous.\cite{SCZhang2DHG,RashbaESHE,DasSarmaESHE} The ISHE in
the 2D hole gas has no vertex correction,\cite{SCZhang2DHG}, and
its existence has been widely accepted,\cite{SHESummary} the
existence of electron ISHE in two-dimensional systems is under
substantial
debate.\cite{InoueHalperinHaldane,MacdonaldALL,RashbaSumRuleRaimondi,EOMArgument,Khaetskii,Nonideal}
The current understanding is that the electron ISHE in the ideal
model (single-band Hamiltonian with parabolic dispersion and
linear Rashba and/or Dresselhaus spin splitting) is exactly
cancelled by the impurity induced vertex corrections in the clean
limit,\cite{InoueHalperinHaldane,MacdonaldALL} even for momentum
dependent
scattering.\cite{RashbaSumRuleRaimondi,EOMArgument,Khaetskii}

Very recently quantum spin Hall effect was predicted theoretically
and observed experimentally in a narrowgap HgTe quantum well with
the unique inverted band structure\cite{Zhang}. The HgTe quantum
well has a quantum phase transition when the quantum well
thickness $d$ is tuned across a critical thickness $d_c\approx 6
nm$. For $d<d_c$, the electronic structure is normal, similar to
the GaAs quantum wells, where the conduction band has $\Gamma_{6}$
character, and the valence band $\Gamma_{8}$ character. In this
regime, we show that the ISHE vanishes in the conduction band due
to vertex corrections, consistent with previous results. For
$d>d_c$, the electronic structure is inverted, where the
conduction and the valence bands interchange their
$\Gamma_{6}-\Gamma_{8}$ characters. In this regime, we show that
the ISHE is finite in the conduction band. Since the impurity
configuration is not expected to change drastically across $d_c$,
the difference of the SHE across the $d_c$ transition therefore
singles out the ISHE contribution. This mechanism solves a long
standing challenge of how to distinguish the ISHE from the ESHE.

First we develop a unified description of $\Gamma_{6}%
$-electron and $\Gamma_{8}$-hole SHE based on a general $N$-band
effective-mass theory, which remains valid over the whole range of $\Gamma
_{6}$-$\Gamma_{8}$ coupling strengths and bandgaps. Then we take the $N$=$8$
model (the Kane model) to make a realistic calculation of the ISHE in
CdTe/Cd$_{x}$Hg$_{1-x}$Te quantum wells, taking into account the non-ideal
factors in a self-consistent way and the impurity scattering induced vertex
corrections through standard diagrammatic techniques. The calculated ISHE
agrees with previous theories in the limit of weak $\Gamma_{6}$-$\Gamma_{8}$
coupling, while it shows nontrivial behaviors in the strong coupling regime.
It exhibits a large [$3\sim4$ times larger than the intrinsic value
$\sigma_{0}=e/(8\pi)$] abrupt increase accompanying the $\Gamma_{6}$%
-$\Gamma_{8}$ phase transition in the lowest conduction band. This large ISHE
is robust against impurity scattering induced vertex corrections. By varying
the well width or the electric bias across the quantum well, we can switch the
electron ISHE on/off or even tune it into resonance. These operations can be
realized in experimentally accessible conditions and they may be utilized to
distinguish the electron ISHE from the ESHE.

Following the new envelope function approach,\cite{Burt} the band-edge Bloch
basis $\{\Phi_{\mu}\}$ is classified into $N$ relevant bands $\{\Phi_{j}\}$
and infinite irrelevant bands $\{\Phi_{l}\}$. In the $N$-dimensional
$\{\Phi_{j}\}$ subspace, the image of the Hamiltonian $H$ for a general
microstructure is $\mathbb{H}_{jj^{\prime}}=\mathcal{H}_{jj^{\prime}}+\sum
_{l}\mathcal{H}_{jl}(E-E_{l})^{-1}\mathcal{H}_{lj^{\prime}}$.\cite{Burt} The
image of an arbitrary operator $O$ $(\neq H$) can also be obtained as
\begin{equation}
\mathbb{O}_{jj^{\prime}}=\mathcal{O}_{jj^{\prime}}+\sum_{l}(\mathcal{O}%
_{jl}\frac{1}{E-E_{l}}\mathcal{H}_{lj^{\prime}}+\mathcal{H}_{jl}\frac
{1}{E-E_{l}}\mathcal{O}_{lj^{\prime}}), \label{ArbOperator}%
\end{equation}
where $\mathcal{H}$ and $\mathcal{O}$ are, respectively, the image of $H$ and
$O$ in the $\{\Phi_{\mu}\}$ space. Then the images of velocity $\mathbf{v}$,
spin $\mathbf{s}$, and spin current $j_{\alpha}^{\beta}\equiv(v_{\alpha
}s_{\beta}+s_{\beta}v_{\alpha})/2\ (\alpha,\beta=x,y,z)$ operators are given
by $\mathbb{\vec{V}}_{jj^{\prime}}=\left[  \mathbf{r},\mathbb{H}_{jj^{\prime}%
}\right]  /(i\hbar)$, $\mathbb{\vec{S}}_{jj^{\prime}}=\left\langle \Phi
_{j}\left\vert \mathbf{s}\right\vert \Phi_{j^{\prime}}\right\rangle $, and
$\mathbb{J}_{\alpha}^{\beta}=(\mathbb{V}_{\alpha}\mathbb{S}_{\beta}%
+\mathbb{S}_{\beta}\mathbb{V}_{\alpha})/2$.\cite{noteOperator} With the
$\Gamma_{6}$-$\Gamma_{8}$ coupling taken into account, they generalize the
previous theories (which neglect this coupling) to the $N$-band case, e.g.,
the widely used single-band (four-band Luttinger-Kohn) model corresponds to
$N$=$2$ $(N$=$4)$. We emphasize that explicit consideration of the $\Gamma
_{6}$-$\Gamma_{8}$ coupling is important in determining electron ISHE,
especially for strong $\Gamma_{6}$-$\Gamma_{8}$ coupled systems. Further, the
different non-ideal band structure factors arise from the same origin
($\Gamma_{6}$-$\Gamma_{8}$ coupling), so they are not independent and should
be incorporated self-consistently through explicit consideration of the
$\Gamma_{6}$-$\Gamma_{8}$ coupling. We notice that the equation-of-motion
argument\cite{EOMArgument} (valid for $N$=$2$) for the nonexistence of
electron ISHE is not applicable to other values of $N$ (e.g., $N$=$4,6$, or
$8$).

\begin{figure}[ptb]
\includegraphics[width=\columnwidth]{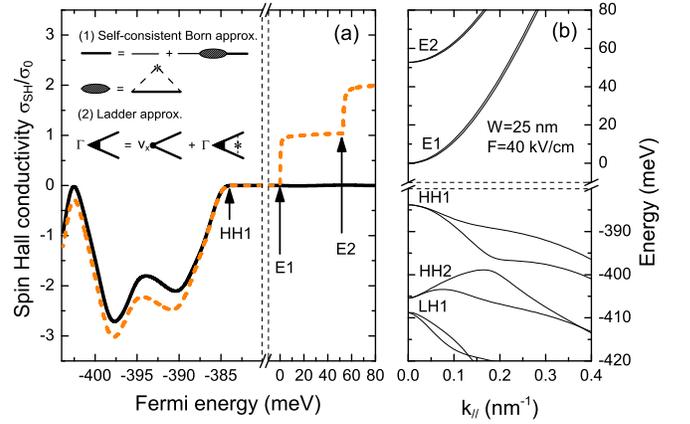}\caption{(color online) (a)
$\sigma_{\text{SH}}$ for $W$=25 nm and $F$=40 kV/cm with (solid lines) or
without (dashed lines) vertex corrections. Inset: (1) Dyson equation in the
self-consistent Born approximation and (2) vertex equation in the ladder
approximation. (b) Corresponding energy spectrum.}%
\label{figP400}%
\end{figure}

The above theory can be applied to study both ISHE and ESHE in a general
microstructure. In the present work we consider ISHE only, due to its much
larger magnitude compared to ESHE,\cite{RashbaESHE,DasSarmaESHE} especially
for small electron density. The linear response spin Hall conductivity
$\sigma_{\text{SH}}=e/(\hbar\mathcal{A})\lim_{\omega\rightarrow0}\left[
G_{xy}^{z}(\omega)-G_{xy}^{z}(0)\right]  /(i\omega)$, with $\mathcal{A}$ the
sample area and $G_{xy}^{z}(\omega)$ the impurity-averaged retarded
correlation function of $\mathbb{J}_{y}^{z}$ and $\mathbb{V}_{x}$. Using
standard diagrammatic perturbation theory, $G_{xy}^{z}(\omega)$ is evaluated
taking into account the impurity induced self-energy corrections in the
self-consistent Born approximation and vertex corrections in the ladder
approximation (inset of Fig. \ref{figP400}), yielding%
\begin{multline*}
\sigma_{\text{SH}}=\frac{e}{\pi}%
{\displaystyle\int\nolimits_{-\infty}^{\infty}}
d\omega\ f(\omega)\lim_{\eta\rightarrow0^{+}}\\
\operatorname{Re}\left[  \frac{\partial P(\omega^{\prime}+i\eta,\omega+i\eta
)}{\partial\omega^{\prime}}-\frac{\partial P(\omega^{\prime}+i\eta
,\omega-i\eta)}{\partial\omega^{\prime}}\right]  _{\omega^{\prime}=\omega},
\end{multline*}
where $f(\omega)=1/[e^{(\hbar\omega-\mu)/(k_{B}T)}+1]$, $P(z,z^{\prime
})=(1/\mathcal{A})\operatorname*{Tr}\mathbb{J}_{y}^{z}\mathcal{G}%
(z)\Gamma(z,z^{\prime})\mathcal{G}(z^{\prime}),$ $z=i\omega_{m},$ $z^{\prime
}=i\omega_{n}$, $\mathcal{G}_{ij}(z)$ and $\Gamma_{ij}(z,z^{\prime})$ are,
respectively, matrix elements of the impurity-averaged Matsubara Green's
function and dressed velocity vertex in the eigenstate basis of $\mathbb{H}$.
They can be calculated from the Dyson equation and the vertex equation%
\[
\Gamma(z,z^{\prime})=\mathbb{V}_{x}+n_{I}%
{\textstyle\int}
d\mathbf{R\ }\frac{\mathcal{U}(\mathbf{R})}{\hbar}\mathcal{G}(z)\Gamma
(z,z^{\prime})\mathcal{G}(z^{\prime})\frac{\mathcal{U}(\mathbf{R})}{\hbar},
\]
where $n_{I}$ is the impurity concentration, and $\mathcal{U}_{ij}%
(\mathbf{R})=\left\langle i\left\vert V_{C}(\mathbf{r}-\mathbf{R})\right\vert
j\right\rangle $ is the matrix element of the single-impurity potential
$V_{C}(\mathbf{r})$.

Now we consider the lattice-matched symmetric CdTe/Cd$_{x}$Hg$_{1-x}$Te
quantum well under electric bias. Its bandgap can be tuned in a large range by
varying the Cd content $x$, the well width $W$, or the bias electric field
$F$, serving as an ideal workbench for studying ISHE under various $\Gamma
_{6}$-$\Gamma_{8}$ coupling strengths. For such narrowgap systems, the $N$=$8$
Kane model is a good starting point. It incorporates the aforementioned
non-ideal factors non-perurbatively and self-consistently. The Dresselhaus
spin-orbit coupling is neglected because it is much smaller than the Rashba
effect in a narrowgap quantum well,\cite{LommerSIA} as verified by the
quantitative agreement between theory and experiment in recent investigations
on the transport properties of CdTe/HgTe quantum wells.\cite{XCZhangHgCdTe} We
also adopt the widely employed axial approximation (good for electrons and
reasonable for holes in narrowgap systems) and short-range impurity potential
$V_{C}(\mathbf{r})=V_{0}\delta(\mathbf{r})$. All band parameters used in our
numerical calculation are experimentally determined
values.\cite{XCZhangHgCdTe,LB} We take the temperature $T$=$0$ K and, unless
specified, the effective disorder strength $\xi(\equiv n_{I}V_{0}^{2})=6.2$
eV$^{2}$ \AA $^{3}$, corresponding to typical electron (hole) self-energy
broadening 0.1 meV (1 meV) and collisional lifetime 6 ps (0.6 ps).

First we consider the weak $\Gamma_{6}$-$\Gamma_{8}$ coupling case $x$=$0.37$
with Eg(Hg$_{0.63}$Cd$_{0.37}$Te)$\approx$0.4 eV (Fig. \ref{figP400}). Without
vertex corrections, the electron ISHE exhibits step-like increases (by
approximately one universal value $\sigma_{0}$) at the edges of the first (E1)
and second (E2) conduction bands. Such behavior is greatly suppressed by the
inclusion of vertex corrections, in sharp contrast to the hole ISHE, which has
an opposite sign and remains largely unaffected by vertex corrections.
Therefore in the weak coupling regime, the results of the previous theories
are recovered.

\begin{figure}[ptb]
\includegraphics[width=\columnwidth]{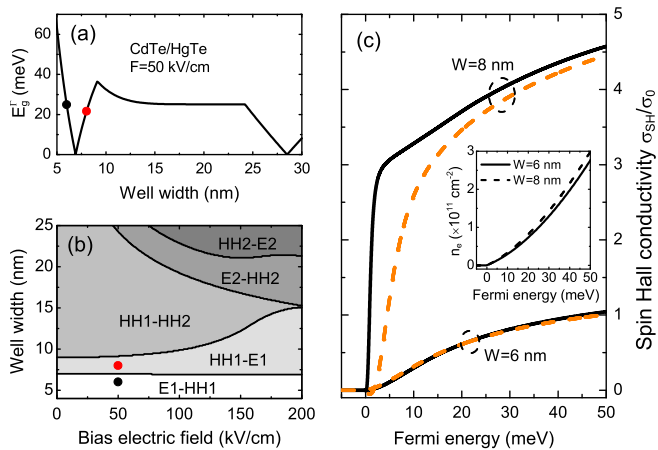}\caption{(color online) (a)
$E_{g}^{\Gamma}$ for $F$=50 kV/cm. (b) Band-edge ($\mathbf{k}_{\parallel}$=0)
phase diagram of the lowest conduction band and highest valence band. (c)
$\sigma_{\text{SH}}$ for $F$=50 kV/cm, $W$=6 and 8 nm, respectively [indicated
by filled circles in (a) and (b)]. Solid (Dashed) lines correspond to $\xi=$
$6.2$ ($62$) eV$^{2}$ \AA $^{3}$. Inset: electron density vs. Fermi energy.}%
\label{figSHCwidth}%
\end{figure}

To explore electron ISHE in the strong coupling regime, we consider the
CdTe/HgTe quantum well corresponding to $x$=$0$. Due to the abnormal positions
and effective masses of the $\Gamma_{6}$ electron and $\Gamma_{7}$ light-hole
bands in the HgTe layer, the bandgap $E_{g}^{\Gamma}$ of the quantum well at
$\mathbf{k}_{\parallel}$=$0$ can be tuned by varying the well width or the
electric bias. Fig. \ref{figSHCwidth}(a) shows that the derivative of
$E_{g}^{\Gamma}$ is discontinuous at $W\approx7,9,24$, and $28.5$ nm,
indicating certain phase transitions. Actually, the first critical point at
$W\approx7$ nm corresponds to the normal-inverted phase transition E1-HH1
$\rightarrow$ HH1-E1.\cite{CVinvert} Namely, the lowest conduction (highest
valence) band changes from E1 to HH1 (HH1 to E1), where E (HH) denote
$\Gamma_{6}$ electron ($\Gamma_{8}$ heavy-hole) states. Other critical points
corresponds to similar transitions [Fig. \ref{figSHCwidth}(b)]. They manifest
the red (blue) shift of electron states E2, E3, $\cdots$ (heavy-hole states
HH1, HH2, $\cdots$) with increasing well width/electric bias due to weakening
of the confinement/quantum confined Stark effect. From Fig. \ref{figSHCwidth}%
(c), we see that in the E1-HH1 phase, $\sigma_{\text{SH}}$ arising from the
lowest conduction band (E1) is largely cancelled by vertex corrections,
especially for small Fermi energy. In contrast, in the HH1-E1 phase, the
lowest conduction band (HH1) takes on pure $\Gamma_{8}$ symmetry at small wave
vectors and its contribution to $\sigma_{\text{SH}}$ is largely
unaffected,\cite{noteMacdonald} leading to the abrupt increase of
$\sigma_{\text{SH}}$ accompanying the phase transition from E1-HH1 to HH1-E1.
This phase transition induced ISHE is robust against impurity induced vertex
corrections since it varies only slightly when $\xi$ is increased by an order
of magnitude, i.e., when typical electron lifetime [mobility] decreases from 6
to 0.6 ps [$3\times10^{5}$ to $3\times10^{4}$ cm$^{2}/($V s$)$]. By changing
the well width, large electron ISHE can be switched on/off, especially for
small Fermi energy or electron density [inset of Fig. \ref{figSHCwidth}(c)].

\begin{figure}[ptb]
\includegraphics[width=\columnwidth]{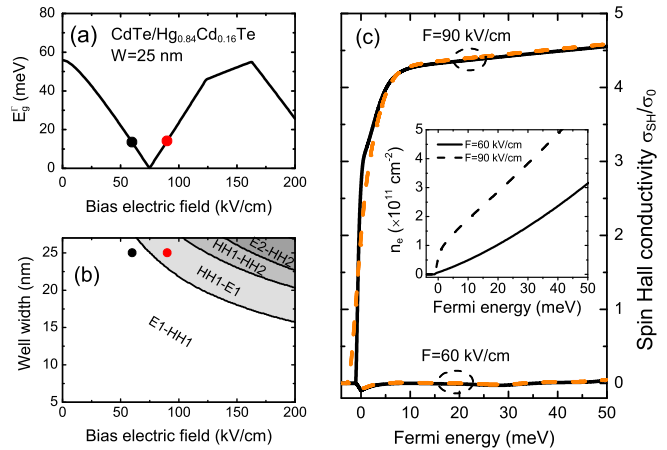}\caption{(color online) (a)
$E_{g}^{\Gamma}$ for $W$=$25$ nm. (b) Band-edge ($\mathbf{k}_{\parallel}$=0)
phase diagram of the lowest conduction band and highest valence band. (c)
$\sigma_{\text{SH}}$ for $W$=25 nm, $F$=60 and 90 kV/cm, respectively
[indicated by filled circles in (a) and (b)]. Solid (Dashed) lines correspond
to $\xi$=$6.2$ ($62$) eV$^{2}$ \AA $^{3}$. Inset: electron density vs. Fermi
energy.}%
\label{figSHCfield}%
\end{figure}

In the above, the phase transition occurs at small critical well width and the
electric bias plays a minor role. When the critical well width increases, the
bias electric field induced quantum-confined Stark effect would become strong
enough to induce the phase transition E1-HH1 $\rightarrow$ HH1-E1 and control
the appearance of large electron ISHE. To demonstrate this, we consider the
case $x$=$0.16$ with Eg(Hg$_{0.84}$Cd$_{0.16}$Te)$\approx$0. For $W$=$25$ nm,
the bandgap $E_{g}^{\Gamma}\approx60$ meV at $F$=0 and decreases to zero at
$F\approx75$ kV/cm [Fig. \ref{figSHCfield}(a)]. The discontinuities of its
derivative at $F\approx75,125$, and $162$ kV/cm clearly manifest the phase
transitions plotted in Fig. \ref{figSHCfield}(b). As a result, $\sigma
_{\text{SH}}$ in Fig. \ref{figSHCfield}(c) shows a large increase when the
bias electric field is tuned across the critical point. Again, the slight
dependence on the disorder strength $\xi$ manifests the robustness of the ISHE
against impurity induced vertex corrections. The field-induced phase
transition provides a dynamic way to switch on/off the electron ISHE,
especially for small Fermi energy or electron density [inset of Fig.
\ref{figSHCfield}(c)].

\begin{figure}[ptb]
\includegraphics[width=\columnwidth]{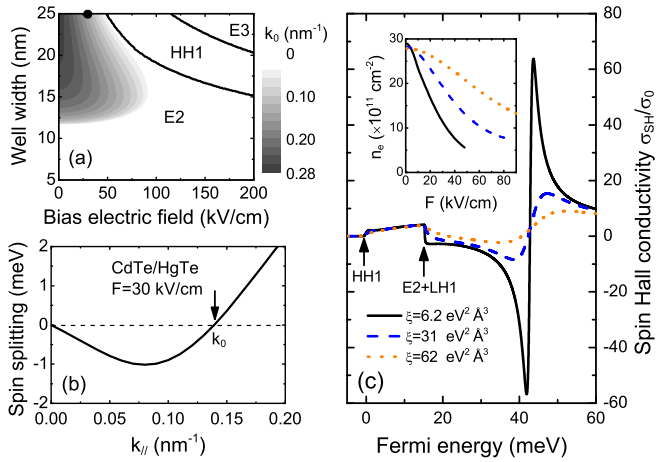}\caption{(color online) (a)
Band-edge ($\mathbf{k}_{\parallel}$=0) phase diagram (the gray scale map in E2
phase indicates $k_{0}$) and (b) Rashba spin splitting (at $W$=25 nm, $F$=30
kV/cm) of the second conduction band. (c) $\sigma_{\text{SH}}$ for $W$=25 nm,
$F$=30 kV/cm [indicated by the filled circle in (a)] and different disorder
strength $\xi$. Inset: critical electron density for $W$=25 (solid line), 20
(dashed line), and 15 (dotted line) nm.}%
\label{figResonance}%
\end{figure}

Turning back to CdTe/HgTe quantum wells, Fig. \ref{figSHCwidth}(a) shows that
the electric bias can induce the transition HH1 $\rightarrow$ E2 in the lowest
conduction band or, equivalently, the transition E2 $\rightarrow$ HH1 in the
second conduction band [Fig. \ref{figResonance}(a)]. In the E2 phase, the
Rashba spin splitting between the two branches of the second conduction band
reverses its sign at a critical wave vector $k_{0}$ [Fig. \ref{figResonance}%
(b)]. Analysis shows that this behavior comes from the coupling between the
two branches and the interface states,\cite{ShamInterfacial} thus it does not
exist in the HH1 phase. By varying the well width or electric bias, such
behavior can be switched on/off [Fig. \ref{figResonance}(a)] and the critical
wave vector [gray scale map in Fig. \ref{figResonance}(a)] or critical
electron density [inset of Fig. \ref{figResonance}(c)] can be tuned, offering
us the possibility to manipulate the ISHE arising from the second conduction
band. Indeed, $\sigma_{\text{SH}}$ in Fig. \ref{figResonance}(c) exhibits a
resonance when the Fermi energy coincides with the spin degeneracy point.

We notice that although such level-crossing induced resonance has
been predicted for the widely accepted hole ISHE in
\textit{p}-type GaAs quantum wells\ (based on calculations that
neglect vertex corrections),\cite{holeISHEResonant} similar
prediction for the much debated \textit{n}-type systems still
remains absent. For hole ISHE, a
challenging hole lifetime $\gtrsim$ 10 ps or hole mobility $\mu_{p}%
\gtrsim10^{4}$ cm$^{2}$/(V s) is required to observe the
resonance.\cite{holeISHEResonant} For electron ISHE, the requirement is
significantly relaxed to electron lifetime $\gtrsim3$ ps [corresponding to
$\xi\lesssim20$ eV$^{2}$ \AA $^{3}$, cf. Fig. \ref{figResonance}(c)] or
electron mobility $\mu_{n}\gtrsim2\times10^{5}$ cm$^{2}$/(V s). These have
already been realized in previous experiments, e.g., $\mu_{n}$=$3.2\times
10^{5}$ cm$^{2}/($V s$)$ for $W$=$7.8$ nm\cite{HgTemobility1} and $\mu_{n}%
$=$3.5\times10^{5}$ cm$^{2}/($V s$)$ for $W$=$21$ nm\cite{XCZhangHgCdTe}
(close to the well width used in our calculation).

In summary, we have investigated the electron ISHE in narrowgap
HgCdTe quantum wells based on a unified description for electron and
hole ISHE. While the ISHE of the conduction band vanishes on the
normal side of the $\Gamma_{6}$-$\Gamma_{8}$ phase transition, a
ISHE in the conduction band can be generated on the inverted side.
It is robust against impurity induced vertex corrections. By
changing the Cd content, the well width, or the bias electric field,
we can switch the ISHE on/off or tune it into resonance under
experimentally accessible conditions. Ref. \cite{Hankiewicz} shows
that the spin Hall effect can be experimentally observed by the
non-local transport measurements in mesoscopic systems. We propose
to carry out such measurement for both the normal and inverted
quantum wells, both close to the transition. The difference uniquely
singles out the ISHE contribution.

This work is supported by the NSFC Grant No. 60525405, the
knowledge innovation project of CAS, the NSF under grant numbers
DMR-0342832 and the US Department of Energy, Office of Basic
Energy Sciences under contract DE-AC03-76SF00515.


\begin{thebibliography}{99}                                                                                               %


\bibitem {Spintronics}S. A. Wolf \textit{et al}., Science \textbf{294}, 1488 (2001).

\bibitem {ESHE}M. I. D'yakonov and V. I. Perel', Phys. Lett. A \textbf{35},
459 (1971); J. E. Hirsch, Phys. Rev. Lett. \textbf{83}, 1834 (1999).

\bibitem {ISHE}S. Murakami, N. Nagaosa, and S. C. Zhang, Science \textbf{301},
1348 (2003); J. Sinova \textit{et al}., Phys. Rev. Lett. \textbf{92}, 126603 (2004).

\bibitem {AwschalomSHE}Y. K. Kato, R. C. Myers, A. C. Gossard, and D. D.
Awschalom, Science \textbf{306}, 1910 (2004); V. Sih \textit{et al}., Nature
Phys. \textbf{1}, 31 (2005); V. Sih \textit{et al.}, Phys. Rev. Lett.
\textbf{97}, 096605 (2006); N. P. Stern \textit{et al}., Phys. Rev. Lett.
\textbf{97}, 126603 (2006).

\bibitem {WunderlichSHE}J. Wunderlich, B. Kaestner, J. Sinova, and T.
Jungwirth, Phys. Rev. Lett. \textbf{94}, 047204 (2005).

\bibitem {SCZhang2DHG}B. A. Bernevig and S. C. Zhang, Phys. Rev. Lett.
\textbf{95}, 016801 (2005).

\bibitem {RashbaESHE}H. A. Engel, B. I. Halperin, and E. I. Rashba, Phys. Rev.
Lett. \textbf{95}, 166605 (2005).

\bibitem {DasSarmaESHE}W. K. Tse and S. Das Sarma, Phys. Rev. Lett.
\textbf{96}, 056601 (2006).

\bibitem {SHESummary}J. Sinova, S. Murakami, S. Q. Shen, and M. S. Choi, Solid
State Commun. \textbf{138}, 214 (2006); and references therein.

\bibitem {InoueHalperinHaldane}E. G. Mishchenko, A. V. Shytov, and B. I.
Halperin, Phys. Rev. Lett. \textbf{93}, 226602 (2004); J. Inoue, G. E. W.
Bauer, and L. W. Molenkamp, Phys. Rev. B, \textbf{67}, 033104 (2003);
\textbf{70}, 041303(R) (2004); D. N. Sheng, L. Sheng, Z. Y. Weng, and F. D. M.
Haldane, \textit{ibid}. \textbf{72}, 153307\ (2005).

\bibitem {MacdonaldALL}K. Nomura, J. Sinova, N. A. Sinitsyn, and A. H.
MacDonald, Phys. Rev. B \textbf{72}, 165316 (2005).

\bibitem {RashbaSumRuleRaimondi}E. I. Rashba, Phys. Rev. B \textbf{70},
201309(R) (2004); R. Raimondi and P. Schwab, \textit{ibid}. \textbf{71},
033311 (2005).

\bibitem {EOMArgument}O. Chalaev and D. Loss, Phys. Rev. B \textbf{71}, 245318
(2005); O. V. Dimitrova, \textit{ibid}. \textbf{71}, 245327 (2005).

\bibitem {Khaetskii}A. Khaetskii, Phys. Rev. Lett. \textbf{96}, 056602 (2006).

\bibitem {Nonideal}A. G. Mal'shukov and K. A. Chao, Phys. Rev. B \textbf{71},
121308(R) (2005); A. V. Shytov, E. G. Mishchenko, H. A. Engel, and B. I.
Halperin, \textit{ibid}. \textbf{73}, 075316\ (2006); C. M. Wang, X. L. Lei,
and S. Y. Liu, \textit{ibid}. \textbf{73}, 113314 (2006); P. L. Krotkov and S.
Das Sarma, \textit{ibid}. \textbf{73}, 195307\ (2006).

\bibitem {Nonparab}D. F. Nelson, R. C. Miller, and D. A. Kleinman, Phys. Rev.
B \textbf{35}, 7770 (1987).

\bibitem {NLRSS}W. Yang and K. Chang, Phys. Rev. B \textbf{73},
113303\ (2006); \textbf{74}, 193314\ (2006).

\bibitem {edgeSHE}\.{I}. Adagideli and G. E. W. Bauer, Phys. Rev. Lett.
\textbf{95}, 256602 (2005).

\bibitem {mesoSHE}L. Sheng, D. N. Sheng, and C. S. Ting, Phys. Rev. Lett.
\textbf{94}, 016602 (2005); W. Ren \textit{et al}., \textit{ibid}.
\textbf{97}, 066603 (2006); Z.\ H. Qiao, W. Ren, J. Wang, and Hong Guo,
\textit{ibid}. \textbf{98}, 196402 (2007).

\bibitem {Zhang}B. A. Bernevig, T. L. Hughes, and S. C. Zhang, Sciences
\textbf{314}, 1757 (2006); M. K\"{o}nig, Science Express 1148047 (2007).

\bibitem {Burt}M. G. Burt, J. Phys.: Condens. Matter \textbf{4}, 6651 (1992).

\bibitem {noteOperator}In obtaining $\mathbb{\vec{S}}$ and $\mathbb{J}%
_{\alpha}^{\beta}$, we have neglected the second term in Eq.
(\ref{ArbOperator}), which is smaller than the first term by a factor
$\mathcal{H}_{jl}/(E-E_{l})\ll1$. This approximation becomes exact when
$\{\Phi_{j}\}$ and $\{\Phi_{l}\}$ are chosen such that $\left\langle \Phi
_{j}\left\vert \mathbf{s}\right\vert \Phi_{l}\right\rangle =0$, e.g., for
$N=2$ (single-band model), $6$ (six-band Luttinger-Kohn model), and $8$ (Kane model).

\bibitem {LommerSIA}G. Lommer, F. Malcher, and U. R\"{o}ssler, Phys. Rev.
Lett. \textbf{60}, 728 (1988).

\bibitem {XCZhangHgCdTe}X. C. Zhang \textit{et al}., Phys. Rev. B \textbf{63},
245305 (2001); K. Ortner \textit{et al}., \textit{ibid}. \textbf{66}, 075322 (2002).

\bibitem {LB}\textit{II-VI and I-VII Compounds; Semimagnetic Compounds},
Landolt-B\"{o}rnstein, Group III, Vol. 41B, ed. U. R\"{o}ssler
(Springer-Verlag, Berlin, 1999).

\bibitem {CVinvert}N. F. Johnson, P. M. Hui, and H. Ehrenreich, Phys. Rev.
Lett. \textbf{61}, 1993 (1988).

\bibitem {noteMacdonald}Based on a modified single-band Rashba model instead
of the correct heavy-hole Hamiltonian, the conclusion of Ref.
\cite{MacdonaldALL} differs from ours. We believe that in the inverted band
phase, single-band results should be taken with cautious due to the strong
$\Gamma_{6}$-$\Gamma_{8}$ coupling.

\bibitem {ShamInterfacial}Y. R. Lin-Liu and L. J. Sham, Phys. Rev. B
\textbf{32}, 5561 (1985).

\bibitem {holeISHEResonant}X. Dai, Z. Fang, Y. G. Yao, and F. C. Zhang, Phys.
Rev. Lett. \textbf{96}, 086802 (2006).

\bibitem {HgTemobility1}J. R. Meyer \textit{et al}., Phys. Rev. B \textbf{38},
2204 (1988).

\bibitem {Hankiewicz}E. M. Hankiewicz, L. W. Molenkamp, T. Jungwirth,
and Jairo Sinova, Phys. Rev. B \textbf{70}, 241301 (2004).
\end{thebibliography}
\end{document}